\documentclass[manuscript]{emulateapj}
\newcommand{\rsun}{\ensuremath{R_\sun}}
\newcommand{\msun}{\ensuremath{M_\sun}}

\newcommand{\kms}{km s$^{-1}$}

\newcommand{\ik}{{\it Kepler} }
\newcommand{\ikp}{{\it Kepler}}

\newcommand{\figr}[1]{Fig.~\ref{fig:#1}}
\newcommand{\secr}[1]{\mbox{Section \ref{sec:#1}}}
\newcommand{\eqr}[1]{Eq.~\ref{eq:#1}}

\newcommand{\tabr}[1]{\mbox{Table~\ref{tab:#1}}}

\shorttitle{Using beaming to measure spin-orbit alignment}

\shortauthors{Shporer et al.}

\begin{document}

\title{On using the beaming effect to measure spin-orbit alignment in stellar binaries with Sun-like components}

\author{Avi Shporer\altaffilmark{1,2}, % LCOGT + UCSB
Tim Brown\altaffilmark{1,2}, % LCOGT + UCSB
Tsevi Mazeh\altaffilmark{3}, % TAU astro
Shay Zucker\altaffilmark{4} % TAU geo
}

% LCOGT (FTN)
\altaffiltext{1}{Las Cumbres Observatory Global Telescope Network, 6740 Cortona Drive, Suite 102, Santa Barbara, CA 93117, USA; ashporer@lcogt.net}
% UCSB (FTN)
\altaffiltext{2}{Department of Physics, Broida Hall, University of California, Santa Barbara, CA 93106, USA}
% TAU astro
\altaffiltext{3}{School of Physics and Astronomy, Tel Aviv University, Tel Aviv 69978, Israel}
% TAU geo
\altaffiltext{4}{Department of Geophysics and Planetary Science, Tel Aviv University, 69978 Tel Aviv, Israel}

%%%%%%%%%%%%%%%%%%%%%%%%%%%%%%%%%%%%%%
\begin{abstract}

The beaming effect (aka Doppler boosting) induces a variation in the observed flux of a luminous object, following its observed radial velocity variation. 
We describe a photometric signal induced by the beaming effect during eclipse of  binary systems, where the stellar components are late type Sun-like stars. The shape of this signal is sensitive to the angle between the eclipsed star's spin axis and the orbital angular momentum axis, thereby allowing its measurement.
We show that during eclipse there are in fact two effects,  
superimposed on the known eclipse light curve. One effect is produced by the rotation of the eclipsed star, and is the photometric analog of the spectroscopic Rossiter-McLaughlin effect, thereby it contains information about the sky-projected spin-orbit angle. The other effect is produced by the varying weighted difference, during eclipse, between the beaming signals of the two stars. 
We give approximated analytic expressions for the amplitudes of the two effects, and present a numerical simulation where we show the light curves for the two effects for various orbital orientations, for a low mass ratio stellar eclipsing binary system. 
We show that although the overall signal is small, it can be detected in the primary eclipse when using \ik Long Cadence data of bright systems accumulated over the mission lifetime.

\end{abstract}
%%%%%%%%%%%%%%%%%%%%%%%%%%%%%%%%%%%%%%

%% Keywords 

%%%%%%%%%%%%%%%%%%%%%%%%%%%%%%%%%%%%%%%%%
\section{Introduction}
\label{sec:intro}
%%%%%%%%%%%%%%%%%%%%%%%%%%%%%%%%%%%%%%%%%

The angle between the spin axis of each star in a binary system and the orbital angular
momentum axis is an important clue for studying stellar binary formation and the effect of tidal processes on  their evolution. Naively one would expect the spins of the two stars and the orbital angular momentum to
be well aligned, as they all originate from the angular momentum of the same primordial
molecular cloud. However, this simple view might be misleading. For example, binary stars
with high obliquity could result from elongated primordial clouds \citep{bonnell92}, or
chaotic accretion during star formation \citep{bate10}. This could also be the result of
interaction with a third body through the Mazeh-Shaham effect \citep{mazeh79},
which is essentially a combination of the so-called Kozai oscillations \citep{kozai62} followed
by tidal dissipation \citep{fabrycky07}. A large sample of binary systems with
measured obliquities is needed for studying the above processes.

To date, spin-orbit alignment has been measured, with spectroscopy, in only three eclipsing stellar binaries, 
two of which are aligned (V1143 Cyg, \citealt{albrecht07}; NY Cep, \citealt{albrecht11}), 
and the third misaligned (DI Her, \citealt{albrecht09}).
It is interesting to note that for transiting planetary systems the spin-orbit angle is
being measured for an increasing number of systems \cite[e.g.,][]{queloz00, winn05, triaud10, winn11}, 
currently including one third of the $\sim$100 known transiting systems. Almost half were identified
as not well aligned, thereby challenging the simplistic understanding of the formation and evolution of planetary
systems (e.g., \citealt{fabrycky09, triaud10, winn10}; see also \citealt{schlaufman10}).

We show here that the beaming effect (aka Doppler boosting; e.g., \citealt{zucker07, mazeh10, shporer10}), combined with the Rossiter-McLaughlin effect (hereafter RM; \citealt{rossiter24, 
mclaughlin24}; see also \citealt{holt1893}), induces a photometric signal during eclipse, where the shape of this signal is sensitive to the spin-orbit angle. Therefore, correctly modeling the light curve allows to measure this angle.
To the best of our knowledge this approach was first used by \cite{hills74}, for a white dwarf. More recently, this approach was noted by \cite{loeb05} for transiting planetary systems, and by \cite{vankerkwijk10} for eclipsing white dwarfs. This signal was theoretically modeled by \cite{groot11} for three kinds of systems: eclipsing double white dwarfs \citep{steinfadt10}, eclipsing O type binaries, and transiting planets orbiting an early type star.
Here we present a further, fourth case: stellar binaries whose components are Sun-like stars, of spectral type F6 and later. Those stars are fundamentally different than earlier type stars and white dwarfs, as they are slow rotaters.

In fact, Sun-like stars are the primary targets of space missions conducting planetary transit surveys like CoRoT and \ikp, where a large number of eclipsing binary systems are monitored. We discuss the possibility of applying this method to bright eclipsing binaries monitored by \ikp, specifically the primary eclipse of low mass ratio binaries, where the low-amplitude signal can be detected by using data obtained during the lifetime of the mission.

We also show that there are in fact {\it two} photometric signals induced by the beaming effect during eclipse, one sensitive to the star's self rotation and spin-orbit angle, while the other is not. Both effects need to be accounted for in order to correctly interpret the light curve and estimate $\lambda$.

We describe the two effects in \secr{effects} and give approximated analytic expressions for their amplitudes. In \secr{sim} we present a numerical simulation of an eclipsing binary composed of an F and K type stars. We discuss the results of the numerical simulation in \secr{dis}, including the possibility of detection with \ik data, and possible degeneracies in the model. We summarize our work in \secr{sum}.

%%%%%%%%%%%%%%%%%%%%%%%%%%%%%%%%%%%%%%%%%
\section{The two beaming-induced signals during eclipse}
\label{sec:effects}
%%%%%%%%%%%%%%%%%%%%%%%%%%%%%%%%%%%%%%%%%

%===============================================================
\subsection{The photometric RM (PRM) effect}
\label{sec:prm}
%===============================================================

The PRM effect results from the combination of two effects, the RM effect and the beaming effect, briefly described in the following paragraphs.

During different phases of an eclipse, the eclipsing star blocks the light coming from different regions of the eclipsed star surface, that have different radial velocities (RVs) due to the rotation of the eclipsed star. The observed spectroscopic line profile is distorted and its center is shifted, resulting in the RM RV signal. The RM RV curve shape depends primarily on the sky-projected angle between the eclipsed star's spin and the system's orbital angular momentum, $\lambda$, the sky-projected rotational velocity of the eclipsed star, $V_{rot}\sin I$, and the secondary to primary radii ratio, $r \equiv R_s/R_p$ \citep[e.g.,][]{ohta05, gimenez06, gaudi07}.

The beaming effect causes the observed flux of a light-emitting object to vary with its RV variation. When the object's RV varies periodically, as is the case in binary systems, it induces a periodic sinusoidal variation in the observed flux. When the light of the secondary can be neglected \citep[e.g.,][]{mazeh10, shporer10}, the amplitude of such a photometric modulation, in relative flux, is  $\alpha_{beam} 4 K/c$, where $K$ is the RV amplitude of the primary and $c$ the speed of light. The coefficient $\alpha_{beam}$ equals unity for bolometric light, and for a finite bandpass its value depends on the target's spectrum and the detector transmission curve. Since in stellar binaries the RV of both components modulates in opposite phase, the observed effect is the weighted difference of the individual beaming effect amplitudes, weighted by their relative fluxes \citep{zucker07}:
\begin{equation}
\label{eq:abeam}
A_{\rm beam} = \frac{4}{c}\frac{\alpha_{beam,p} K_p F_p-  \alpha_{beam,s} K_s F_s}{F_p+F_s} \ ,
\end{equation}
where $\alpha_{beam,p}$, $K_p$, and $F_p$ are the primary's beaming coefficient, RV amplitude and flux, respectively, while $\alpha_{beam,s}$, $K_s$, and $F_s$ are the corresponding quantities for the secondary.
Several authors have already observed this effect, both from space \cite[e.g.,][]{vankerkwijk10, mazeh10, bloemen11, carter11} and from the ground (\citealt{shporer10}, see also \citealt{maxted00}).

When we apply the beaming effect to the eclipsed star rotational velocity, we get the photometric analog of the RM effect which we refer to as the photometric RM effect, or PRM. Photometrically, the blue shifted photons emitted from the star's hemisphere rotating towards the observer make that hemisphere appear slightly brighter than the other, redshifted hemisphere, rotating away from the observer. It is this apparent non uniform surface brightness that causes a photometric anomalous signal during eclipse. 

We adopt a coordinate system centered at the eclipsed star, and the Y axis directed towards the observer while the Y-Z plane contains the eclipsed star rotation axis. Therefore, the variation in relative flux during primary eclipse does not depend on the secondary position along the Y axis, but only on the position within the X-Z plane, and can be described by:
\begin{equation}
\label{eq:aprmint}
\Delta f = \frac{4 \alpha_{beam}}{c} {\cal D} \Omega \sin I \frac{\int\int{x {\cal I}(x,z) dx dz}}{\int\int{ {\cal I}(x,z) dx dz}} \ ,
\end{equation}
where $\Omega$ and ${\cal I}(x,z)$ are the eclipsed star rotation rate and surface brightness (not considering the PRM effect), respectively, and ${\cal D}$ is the dilution factor, accounting for light from the eclipsing star, so $\Delta f$ is the observed relative flux difference. The integration includes the non eclipsed stellar surface and we assume flux of unity when the stars are out of eclipse.
\eqr{aprmint} is similar to the corresponding equation for the spectroscopic RM effect \citep[e.g.,][]{ohta05}, where the multiplication between $\Omega$ and $x$ gives the RV at position ($x$,$z$).

An order of magnitude estimate of the PRM amplitude is:
\begin{equation}
\label{eq:aprm}
A_{\rm PRM} \approx  4 \alpha_{beam} \frac{V_{rot} \sin I}{c} \frac{r^2}{1-r^2} \approx  10^{-5} \frac{V_{rot} \sin I}{10 {\rm \ km \ s^{-1}}} \frac{r^2}{0.1}\ .
\end{equation}
We caution that \eqr{aprm} is only a rough estimate, and $A_{\rm PRM}$ decreases for orbits with an inclination angle of $i\ \textless\ 90$ deg. The primary assumptions behind \eqr{aprm} are that the eclipsing star is relatively small and faint compared to the eclipsed star.

%===============================================================
\subsection{The in-eclipse orbital beaming (InOrB) effect}
\label{sec:inorb}
%===============================================================

The orbital beaming effect follows the Keplerian motion of the  two stars. Therefore, at the time of conjunctions the orbital beaming modulation is an almost completely linear variation with time, especially for a circular orbit. When an eclipse occurs the observed fractional brightness of the two stars changes, leading to a variation in the weights in \eqr{abeam}, resulting in a deviation from the orbital beaming light curve. 
We refer to this deviation as the in-eclipse orbital beaming, or InOrB. 

The amplitude of this second effect depends on the variation of the eclipsed star's relative brightness, and its RV variation during eclipse. The latter can be approximated in the following way, for a circular orbit. The RV amplitude of a star of mass $M_1$ with a binary companion of mass $M_2$ is:
\begin{equation}
\label{eq:k}
K_1 = \frac{2 \pi a}{P_{orb}}\frac{M_2 \sin i}{M_1 + M_2} \ ,
\end{equation}
where $a$ is the orbital semi major axis, $P_{orb}$ the orbital period, and $i$ is the orbital inclination angle. The orbital phase at the time of eclipse start relative to the mid eclipse time is close to $\Delta \phi \approx (R_1+R_2) \sqrt{1-b^2} / a$, where $R_1$ and $R_2$ are the radii of the two stars, and $b$ is the eclipse impact parameter\footnote{defined as $b=\frac{a \cos i}{R_1+R_2}$}. In addition, we assume the orbital phase covered during eclipse is a small part of the orbit, which results in:
\begin{equation}
\label{eq:ainorb}
\Delta RV_{\rm eclipse} \approx K_1 \Delta \phi \approx \frac{2\pi(R_1+R_2)}{P_{orb}}\frac{M_2 \sin i} {M_1+M_2}\sqrt{1-b^2} \ ,
\end{equation}
where $M_1$ and $R_1$ are the mass and radius of the eclipsed star, respectively, and $M_2$ and $R_2$ are the mass and radius of the eclipsing star, respectively.

A rough estimate of the two effects amplitudes ratio gives:
\begin{equation}
\label{eq:ratio}
\frac{A_{\rm InOrB}}{A_{PRM}} \approx \frac{P_{rot,1}}{P_{orb}}\frac{M_2}{M_1+M_2}\frac{R_1+R_2}{R_1}\frac{\sin i}{\sin I_1} \ ,
\end{equation}
where $P_{rot,1}$ is the eclipsed star rotation period. \eqr{ratio} shows that for certain systems the two effects can have similar amplitudes, especially for binaries of similar components where the stars' rotation is synchronized with the orbital period.

\begin{figure*}
\begin{center}
\includegraphics[scale=0.49]{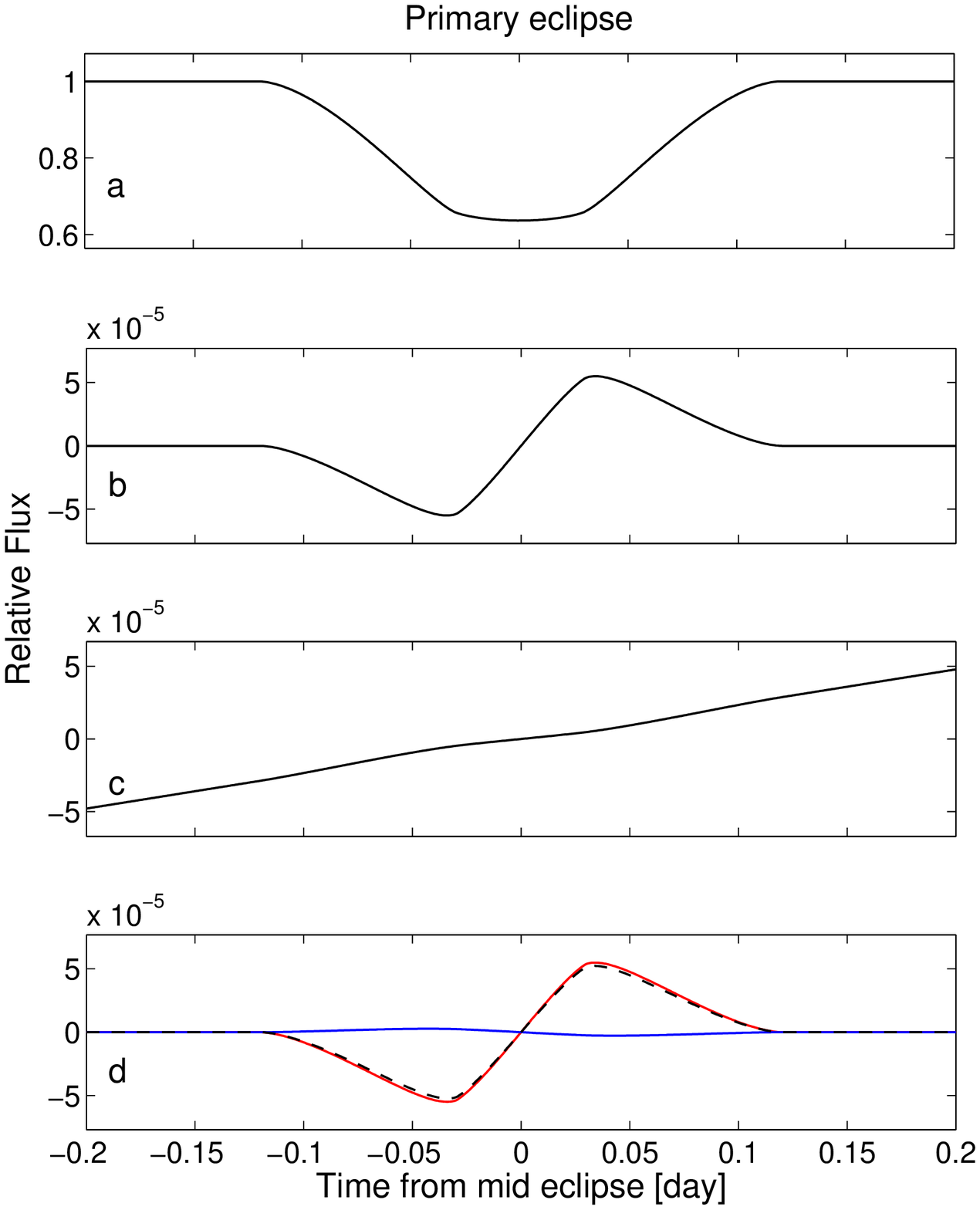}
\hspace{-9mm}
\includegraphics[scale=0.49]{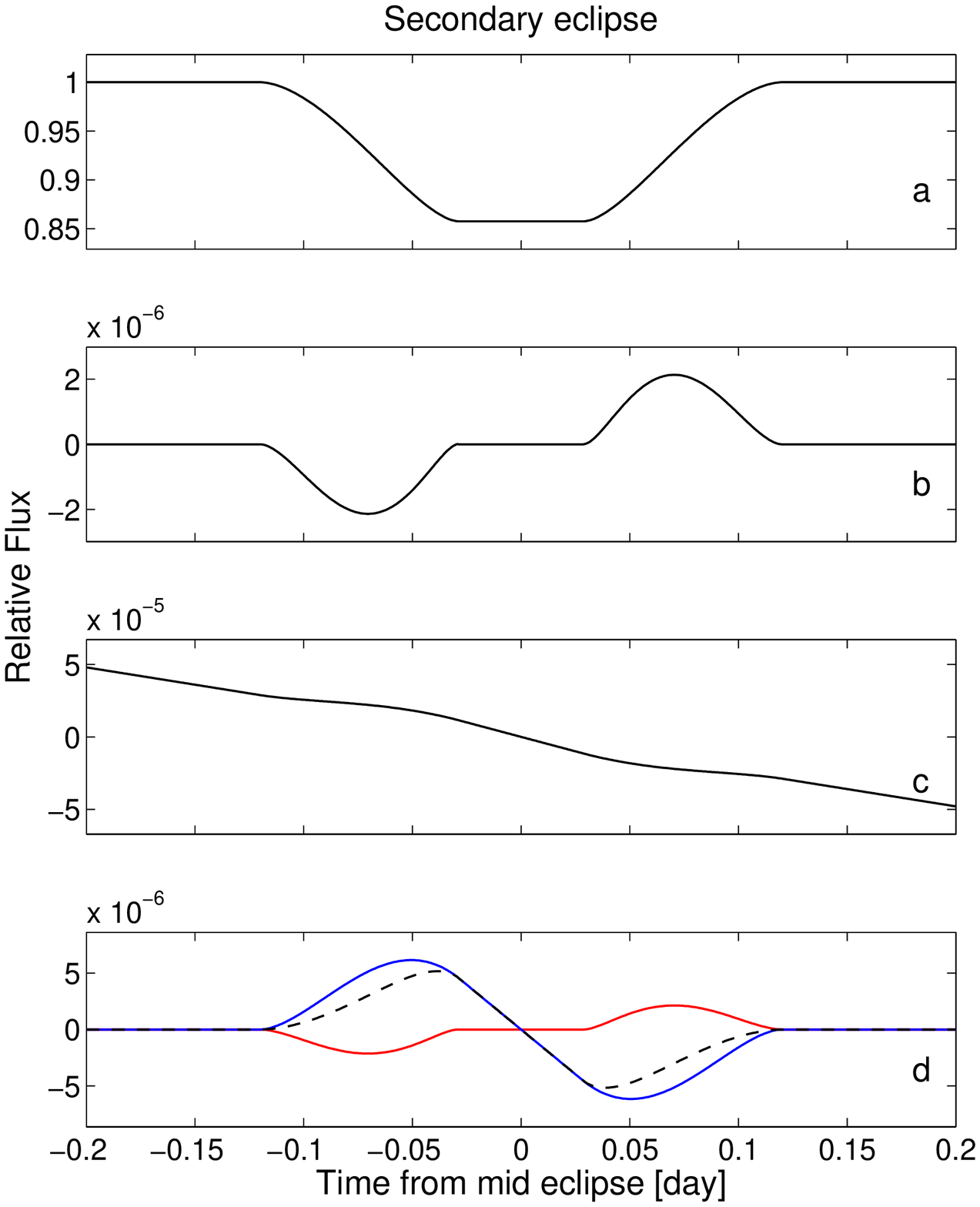}
\caption{\label{fig:F8K0} 
Photometric variability during primary (left) and secondary (right) eclipses of the F8V+K0V eclipsing binary whose parameters are listed in \tabr{params}. In this figure we assumed an edge-on orbit ($i$=90 deg) and spin-orbit alignment ($\lambda$=0 deg) for both stars. From top to bottom: (a) eclipse light curve, accounting for geometry and limb darkening, in relative flux. (b) The PRM light curve. (c) The orbital beaming light curve, including the in-eclipse beaming. (d) The two beaming-induced effects in eclipse, PRM in red and the InOrB in blue. The dashed black line marks their sum. \tabr{params} lists the parameters of the system analyzed here. As expected, during secondary eclipse the PRM effect is much smaller, since the secondary rotates more slowly and is fainter than the primary, and it occurs only during eclipse (occultation) ingress and egress. In fact, the PRM amplitude in the secondary eclipse is smaller than the InOrB amplitude.
}
\end{center}
\end{figure*}

\begin{figure*}
\begin{center}
\includegraphics[scale=0.56]{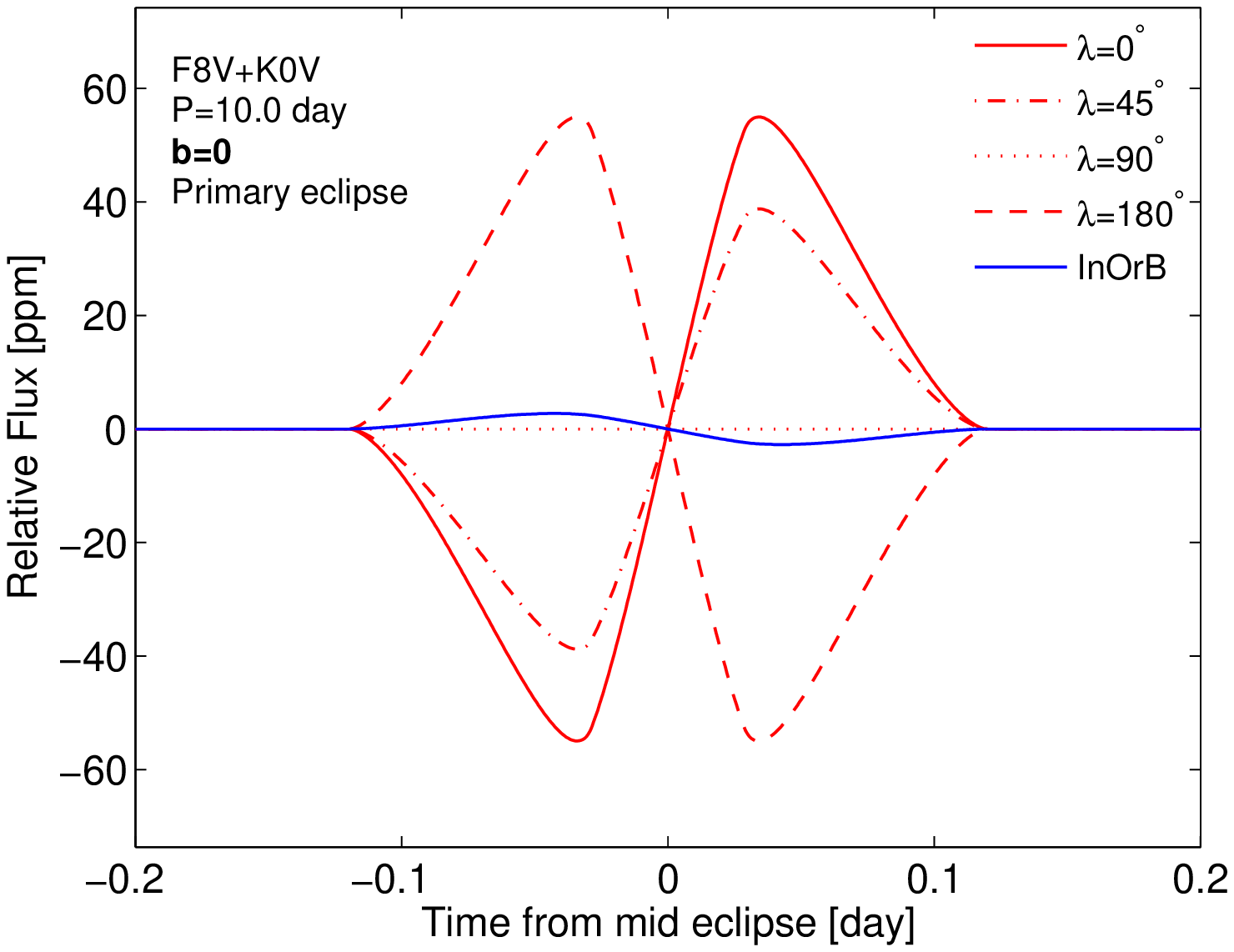}
\hspace{-9mm}
\includegraphics[scale=0.56]{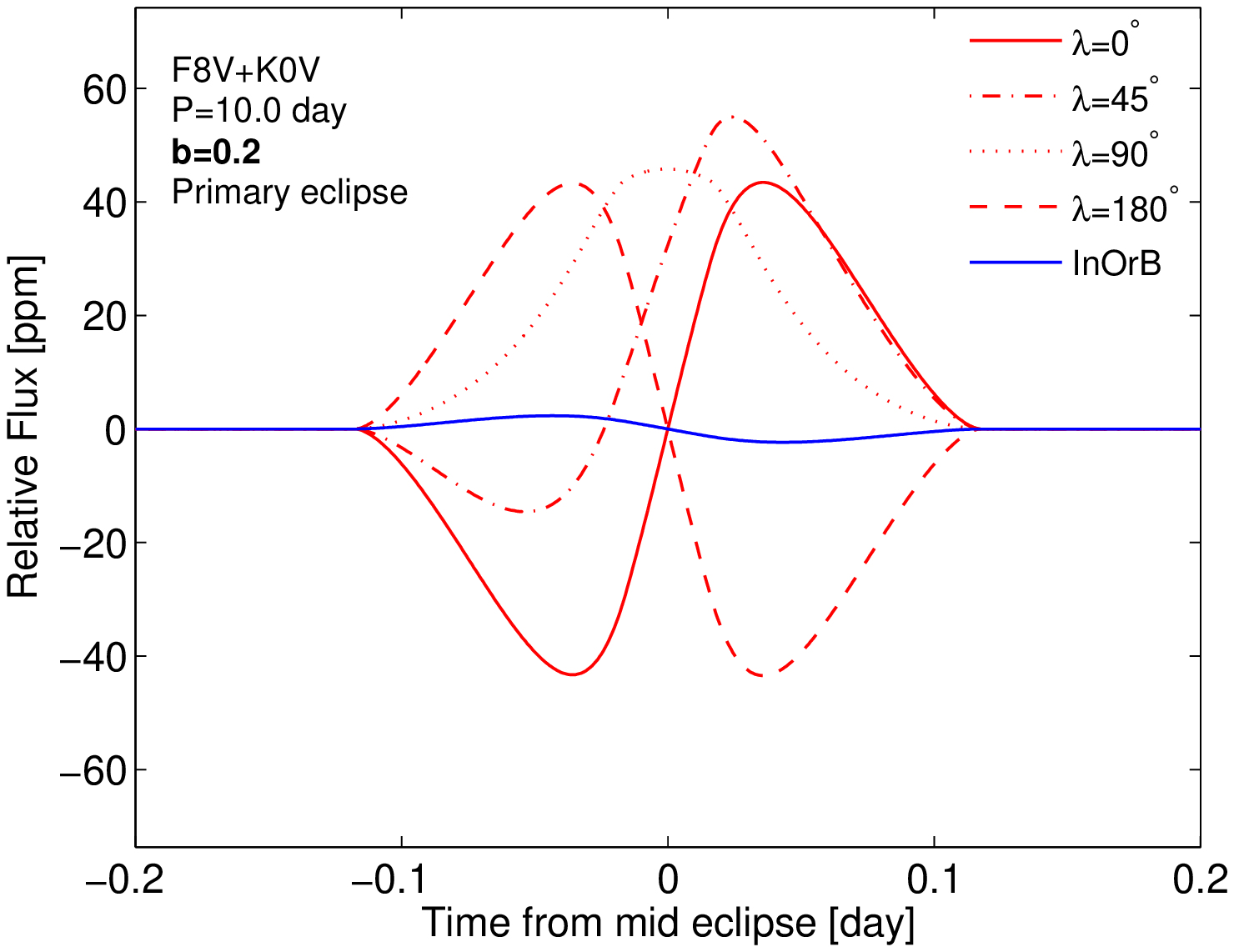}
\caption{\label{fig:lambda} 
Comparison of the beaming-induced signals during primary eclipse, for different impact parameters and spin-orbit angles. The left panel uses $b$=0 ($i$=90 deg), and the right panel $b$=0.2 ($i$=89.14 deg). In both panels the red lines represent four different values of $\lambda$, listed in the panels' legends. The blue solid line shows the InOrB light curve, which does not depend on $\lambda$.
}
\end{center}
\end{figure*}

As \figr{F8K0} and \figr{lambda} show, for prograde orbits the InOrB signal is opposite in sign to the PRM signal, and has a somewhat different shape, as the PRM signal at different phases of the eclipse depends on the sky-projected distance of the eclipsing object from the eclipsed object's rotation axis, while the InOrB effect does not. In fact, the InOrB effect does not depend on the eclipsed star rotation, and occurs even if the star did not rotate at all. As it does not hold information about spin-orbit alignment, or any other new information, the InOrB effect is determined by parameters known from the eclipse light curve (considering only geometry and limb darkening) and the out-of-eclipse (i.e. orbital beaming) light curve. 

%%%%%%%%%%%%%%%%%%%%%%%%%%%%%%%%%%%%%%%%%
\section{Numerical simulation}
\label{sec:sim}
%%%%%%%%%%%%%%%%%%%%%%%%%%%%%%%%%%%%%%%%%

To show the shape and amplitude of the light curves induced by the effects described in \secr{effects} we examine a hypothetical eclipsing binary system and construct the light curves using numerical integration across the observed stellar surfaces. Our numerical code is based on the code used in \citet[][presenting a low-amplitude effect related to the spectroscopic RM effect]{shporer11}.

Our code assumes the two stars are rotating spherical bodies, orbiting each other in a Keplerian orbit, and radiating as blackbodies with given effective temperatures. We further assume their surface brightness obeys a quadratic limb darkening law and we use limb darkening coefficients based on the tabulation of \cite{claret11}, calculated specifically for the \ik transmission curve. For simplicity we use a circular orbit. The beaming coefficient for each star was estimated by integrating the relevant blackbody spectrum across the \ik transmission curve.

The binary system we adopt here is composed of an F8V primary, with $T_{eff,p}$=6250 K, and a K0V secondary, with $T_{eff,s}$=5150 K, orbiting each other every 10.0 days. We assumed solar metallically and an age of 1 Gyr, and derived the stars' masses and radii using the Y$^2$ models \citep{yi01}. We use $M_p=1.20\ M_{\sun}$ and $R_p=1.15\ R_{\sun}$ for the primary, and $M_s=0.80\ M_{\sun}$ and $R_s=0.70\ R_{\sun}$ for the secondary. For the rotation rates we use $V_{rot,p}=20$ \kms\ and $V_{rot,s}=5$ \kms\ for the primary and secondary, respectively, based on \cite{meibom11}. All system parameters are listed in \tabr{params}. These parameters and the above assumptions imply an orbital semi major axis of 0.11 AU.

\begin{deluxetable}{lcc}
\tablecaption{\label{tab:params}
Parameters of the system used in the examples presented in Figs.~1 and 2.} 
\tablewidth{0pt}
\tablehead{\colhead{Parameter} & \colhead{Value} & \colhead{Units}  } 
\startdata
\multicolumn{3}{l}{Orbital parameters:} \\
Orbital Period, $P_{orb}$ 					& 10.0 	& day  \\
Eccentricity, $e$					& 0		& --- \\
\hline
\multicolumn{3}{l}{primary stellar parameters:} \\
Mass, $M_p$						& 1.20	& \msun  \\
Radius, $R_p$						& 1.15	& \rsun  \\
Temperature, $T_{eff,p}$				& 6250	& K \\
Rotational velocity, $V_{rot,p}$			& 20		& \kms  \\
\hline
\multicolumn{3}{l}{primary stellar parameters:} \\
Mass, $M_s$						& 0.80	& \msun  \\
Radius, $R_s$						& 0.70	& \rsun  \\
Temperature, $T_{eff,s}$				& 5150	& K \\
Rotational velocity, $V_{rot,s}$			& 5		& \kms  \\
\enddata
\end{deluxetable}

\figr{F8K0} presents the various effects occurring during primary (left panel) and secondary (right panel) eclipses, where we have assumed an edge-on orbit (inclination angle of $i$=90 deg, or an impact parameter of $b$=0), and complete spin-orbit alignment for both stars ($\lambda_p = \lambda_s = 0$). The eclipse duration for this system is 5.73 hour. The top row panels of \figr{F8K0} show the familiar eclipses light curves, in relative flux versus time, according to the geometry of the system and the limb darkening of the stellar surfaces. The effects induced by  beaming are too small to be visually identified at the scale presented in the top panels, so they are presented in the three bottom rows. The second to top row panels show the PRM light curve. For a prograde system, the approaching side of the star is occulted first. Due to the beaming effect this side is slightly brighter than the receding side, leading to a small flux decrease during the first half of the eclipse, and a flux increase during the second half. The schematic shape of the PRM light curve is similar to the spectroscopic RM RV curve for stellar binaries \citep[e.g.,][]{albrecht07} and transiting planets \citep[e.g.,][]{gaudi07}. For the secondary eclipse, the PRM amplitude is much smaller than for the primary eclipse, since the secondary is fainter, smaller, and rotates more slowly than the primary. Also, the PRM effect is seen only during ingress and egress of the secondary eclipse (whose duration is 2.18 hours) since the star is completely occulted during the 1.37 hours of the full eclipse phase. The second to bottom row panels show the orbital beaming around the time of superior and inferior conjunctions. The linear trend seen in both panels is the orbital beaming, following the systems Keplerian motion. The small deviation from a straight line is the InOrB effect, described in \secr{inorb}. Finally, the bottom row panels show the two beaming-induced effects during eclipse, the PRM in red solid lines and the InOrB in blue solid lines where the linear orbital trend is subtracted, so these panels show only the anomalous anti-symmetric signals that result from the eclipsing nature of the system. The dashed black line is the sum of the two effects, for reference. As can be seen in the bottom left panel, during primary eclipse the PRM amplitude is much larger than the InOrB amplitude. However, this relation is reversed in the secondary eclipse. 

In \figr{lambda} we focus on the primary eclipse and show the beaming-induced signals for different eclipse impact parameters and spin-orbit orientations. The left panel shows the same system as in \figr{F8K0} and \tabr{params}, but for four different values of $\lambda_1$. The red lines show the PRM light curve for spin-orbit angles of 0 (solid line), 45 (dot-dashed line), 90 (dotted line) and 180 deg (dashed line). The figure shows how the PRM signal decreases in amplitude but remains antisymmetric for this completely edge-on case. The blue solid line shows the InOrB effect, which is insensitive to spin-orbit angle, so it is the same for all four orientations. The right panel presents a similar system, only with an impact parameter of $b$=0.2, which is equivalent to $i$=89.14 deg, and it shows how the PRM light curve becomes distorted for non-aligned and non-edge-on systems.
 
%%%%%%%%%%%%%%%%%%%%%%%%%%%%%%%%%%%%%%%%%
\section{Discussion}
\label{sec:dis}
%%%%%%%%%%%%%%%%%%%%%%%%%%%%%%%%%%%%%%%%%

For the edge-on and aligned system presented in \figr{F8K0}, the PRM amplitude is 55 ppm for the primary and 2.2 ppm for the secondary eclipses. These amplitudes are smaller than predicted by \eqr{aprm} due to dilution from the eclipsing star and also because that equation was derived by assuming the eclipsing object is much smaller than the eclipsed one. When this is not the case the integration across a significant fraction of the eclipsed stars' surface acts to decrease $A_{PRM}$ by averaging down the variation in the radial component of the rotational velocity. The InOrB effect amplitude for the primary is 2.7 ppm, which is 5\% of the PRM amplitude, not far from the 0.1 ratio predicted by \eqr{ratio}. For the secondary eclipse the numerically measured InOrB amplitude is 6.2 ppm, almost 3 times larger than the PRM amplitude, much larger than the ratio of 0.6 predicted by \eqr{ratio}. This is not surprising since the assumptions done when deriving \eqr{ratio} are invalid for the secondary eclipse.

Both beaming-induced eclipse signals presented here are small, and can be detected only with ultra precise photometry. To estimate whether they can be detected by \ik we assume an observed brightness of 12th magnitude, where according to the \ik Mission specifications \citep[e.g.,][]{borucki10, koch10} the photometric precision is 20 ppm over 6.5 hours (to allow detection of the 13 hours long and 84 ppm deep transit of an Earth-like planet across a Sun-like star). Assuming white noise this translates to 72 ppm per 0.5 hour. Such a precision will of course not allow a detection from a single eclipse event even for the primary eclipse. Fortunately, \ikp's long lifetime, of 3.5 years, will allow to fold together many eclipses and decrease the scatter by $\sqrt{N}$ where $N$ is the number of observed eclipse events. For the 10 day system adopted here it is fair to assume that 100 eclipse events will be observed, even when accounting for gaps following data download time or  spacecraft Safe Modes. Therefore, an accuracy of about 7 ppm per half an hour is predicted to be achieved over the lifetime of the mission. Such a precision should be sufficient to detect the PRM effect during primary eclipse for a system similar to the one simulated here.

Most of the \ik targets are observed in the default Long Cadence mode. In this mode every 270 individual consecutive exposures are stacked on-board, each of 6.02 s exposure time and 0.52 s readout time. This results in an effective cycle time of 29.4 min where photons are collected during 91.4\% of that time. The high duty cycle allows us to treat each cycle as half an hour exposure.
Hence, the individual eclipse events of the system shown in \figr{F8K0} are sampled with 12--13 exposures, which is sufficient for resolving the temporal flux variation although the variation across half an hour is averaged out. The smaller PRM signal during secondary eclipse makes it more difficult to detect, and probably not detectable with \ikp, although the amplitude increases for stars rotating more rapidly, which is expected for younger stars due to the Skumanich spin-down law \citep{skumanich72}.

The system we examined here has stellar components which are significantly different from each other. If we take them to be more similar, then the secondary eclipse signal becomes larger, although the primary smaller. For a system composed of two G0V stars with a self rotation velocity of 10 \kms, our numerical calculations show that the PRM amplitude will be close to 15 ppm. For such a system the InOrB amplitude will be 11 ppm, larger than for the case of the F8V+K0V system.

The signal to noise ratio of the in-eclipse light curve is determined by the total amount of time the system spends in eclipse. Since for given stellar radii this scales with $P^{-2/3}$, eclipsing systems orbiting with a shorter orbital period will require data taken over a shorter time span in order to reach the required precision. Taking the same 12th magnitude system adopted here, only with $P_{orb}$=1.5 day gives a precision of 7 ppm per half an hour after 150 days of data, within two \ik quarters\footnote{\ik quarters have a duration of about 90 days.}. However, the eclipse duration of such a short period system is only 3.07 hours. This means that while $A_{PRM}$ remains the same, each eclipse will be covered by 6--7 Long Cadence exposures, and each half of the eclipse by only 3--4 exposures. Moreover, when all Long Cadence measurements are folded together and binned the effective averaging of the flux variability goes beyond half an hour. Therefore, using Long Cadence data to detect the PRM signal of short period eclipsing binaries will be difficult, and even more so for distinguishing between the different possible spin-orbit orientations. Using \ik Short Cadence data, with an effective integration time of 1 minute (9 readouts), could make it possible to detect the PRM effect in short period binaries. However, the Short Cadence mode is applied only to a small sample of 512 objects, about 0.3\% of the \ik targets, which are carefully chosen and observed in this mode usually during only one \ik quarter.

Measuring spin-orbit alignment for longer period systems, with a period of 10 days, as simulated here, is also expected to be more useful to study binary formation, compared to short period systems. Processes resulting from tidal interaction between the two stars, including orbital circularization, synchronization between the orbital period and the stars'  rotation, and spin-orbit alignment \citep{mazeh08}, have a time scale that is strongly dependent on the scaled semi-major axis, $a/R_p$ \cite[e.g.,][]{zahn89}. The spin-orbit alignment specifically has a relatively short time scale, two or three orders of magnitude shorter than the circularization time scale since it involves a transfer of a smaller amount of angular momentum \citep{mazeh08}. Therefore, it is expected that for short period binaries the spin axes of the two stars had already aligned with the orbital momentum axis, and any initial missalignment --- a relic from their formation history --- had already faded away. Since for longer period systems, with a wider orbit, this process is slower, they are better targets for studying the initial alignment and the systems' formation process. For example, the missaligned system DI Her (\citealt{albrecht09}, see also \citealt{mazeh08}) has an orbital period of 10.6 day. 

We used the \ik eclipsing binary catalog of \cite{prsa11}, including 1879 systems, to estimate the number of eclipsing binary systems whose photometry will be accurate enough to allow looking for the PRM signal. We looked for systems whose eclipse duration is longer than 5 hours, and that given their orbital period and brightness (\ik magnitude) the photometric S/N will be no less than half that of the 10 days and 12th mag system examined here. This includes 10 day systems whose brightness is down to 13.5 mag, or 12 mag systems with period up to 40 days.
This cut left about 320 systems. When we restrict ourselves to low mass ratio systems, similar to the one we examined here, this number is about 50. A detection of the PRM effect and a measurement of the spin-orbit angle for the primary in even some of those systems will greatly extend the sample of eclipsing binaries for which this angle is known.

% ellip and refl
In our numerical analysis we ignored several low-amplitude effects, including ellipsoidal distortion due to tidal forces between the two stars and reflection of light, or heating of one star by the other. For a 10 day period binary both these effects are expected to have an amplitude along the orbit which is smaller than that of the orbital beaming effect \citep{zucker07}. In addition, during eclipses the ellipsoidal and reflection effects are at extrema, meaning their induced flux variation is minimal, while the orbital beaming flux is at its phase of fastest variation. The light curves of the ellipsoidal and reflection effects during eclipse are expected to be symmetric about the mid eclipse time, and, as for the orbital beaming, they can be predicted using modeling of the out of eclipse light curve \citep{faigler11}. 

% activity
We note that we ignore here the possibility of other sources of low level noise, such as stellar noise, meaning that this measurement requires quiet stars. In case the eclipsed star shows slight activity in the form of small spots on its surface, this can lead to sinusoidal-like flux variability at the rotational period. However, for the type of system simulated here the orbital and self rotation periods are not necessarily synchronized, so the spots-induced flux variability will be averaged out once folded on the orbital period, which is determined to high accuracy by the measurement of eclipses timing.

% ecc
Our method has several possible degeneracies, which if not carefully considered can lead to incorrect interpretation of the light curve, and a biased value for $\lambda$. Significant eccentricity can produce asymmetric eclipse light curves, depending on the argument of periastron, $\omega$ (see \citealt{kipping08} for a description of the impact of eccentricity on planetary transit light curves). Although, eccentricity can be estimated from the orbital beaming light curve, and, the value of $e\cos \omega$ is accurately determined from the time difference between the primary and secondary eclipses. 

% mid transit time shift
A second degeneracy results from the PRM signal being mimicked by a small shift in mid eclipse time, which may give the false impression that the PRM signal does not exist in the light curve, suggesting a scenario where $i$ = 90 deg and $\lambda$ = 90 deg, shown in \figr{lambda} left panel. We looked into this possibility more closely by subtracting the eclipse light curve including a small shift in mid eclipse time, and without the PRM effect, from the light curve including the PRM signal and no time shift. The mid eclipse time shifts we examined were up to a few seconds, in sub second steps. We found that the peak to peak amplitudes of the difference light curves were at least 60 ppm, and the smallest difference is for time shifts close to 1 s. We expect \ik photometry obtained during the time span discussed here to allow mid eclipse measurements of better than 1 s precision. Therefore, a solution which does not include the PRM effect  (or assumes $i$ = 90 deg and $\lambda$ = 90 deg) and as a result gives a wrong mid eclipse time due to the above mentioned degeneracy will have a significantly lower likelihood than a solution that does include the PRM effect.

Modeling the PRM effect adds two parameters to the model, $V_{rot}\sin I$ and $\lambda$ of the eclipsed star. An independent estimate of $V_{rot}\sin I$ can be achieved from a spectrum of the star, thereby constraining this parameter so the information in the PRM signal is used for fitting only a single parameter ($\lambda$). Moreover, an independent estimate of $V_{rot}\sin I$ is important for lifting a third degeneracy in our model, between $\lambda$ and $V_{rot}\sin I$, for edge-on orbits \citep[e.g.,][]{gaudi07}.

Another possible degeneracy may result from the InOrB signal being opposite in sign to the PRM signal (see Figures~\ref{fig:F8K0} and \ref{fig:lambda}), giving a false impression that the PRM signal is smaller in amplitude. In general, for non-aligned and non edge-on orientations, the combined light curve shape could be different than the PRM light curve, possibly leading to a biased estimate of $\lambda$, even though for the primary eclipse shown in Figures~\ref{fig:F8K0} and \ref{fig:lambda} the PRM signal is clearly dominant. However, as already noted, the InOrB light curve shape is insensitive to the spin-orbit angle or the star's rotation velocity. The only parameter to which both signals are sensitive to is the radii ratio $r$, which is accurately determined from the eclipse light curve even without accounting for the beaming-induced effects. Therefore, the InOrB light curve is determined independently of the PRM light curve, eliminating this possible degeneracy.

% self rotation
Lastly, we note that the oblateness of stars, due to self rotation, may also affect eclipse light curves, especially when oblateness leads to significant gravity darkening. \cite{barnes09} showed how this will distort the shape of planetary transit light curves when the host star is a rapidly rotating early type star. However, since Sun-like stars are slow rotators this is expected to have a minor impact on the light curves (for the Sun for example the measured oblateness\footnote{The fractional difference in the radius at the equator and the poles.} is 10$^{-5}$).

% planets
It is interesting to consider the application of the method presented here to a transiting planet orbiting a Sun-like star. However, as seen from \eqr{aprm}, in this case $A_{PRM}$ is expected to be at the 1 ppm level, making it difficult, if at all possible to detect. The case of a transiting planet orbiting an A type star is presented by \cite{groot11}, where the assumed rotational velocities are at the order of 100 \kms, and the presented light curve has an amplitude of about 10 ppm.

%%%%%%%%%%%%%%%%%%%%%%%%%%%%%%%%%%%%%%%%%
\section{Summary}
\label{sec:sum}
%%%%%%%%%%%%%%%%%%%%%%%%%%%%%%%%%%%%%%%%%

We have presented the effects induced by beaming during eclipses of binary systems composed of Sun-like stars. Although a detection of these effects is clearly highly challenging, we have presented a numerical simulation of a F8V + K0V eclipsing binary where data accumulated over the \ik Mission lifetime should allow to detect it during primary eclipse, and in turn measure the spin-orbit alignment of the primary. Given the known characteristic of \ik eclipsing binaries we expect there will be about 50 systems similar to the system simulated here, where the PRM signal can be measured. In case the \ik mission will be extended, this number will become significantly larger. So far the spin-orbit angle was measured in only a few stellar binary systems, making its measurement in any additional system of high scientific value.

The method presented here complements other methods used for obliquity measurement in stellar binaries and star-planet systems. Those include the measurement of the spectroscopic RM effect \citep{albrecht07, albrecht09, albrecht11, winn05, triaud10}, and the photometric methods presented by \cite{barnes09}, for rapidly rotating early type planet host stars, and by \cite{sanchis11}, for active planet host stars. In addition, \cite{groot11} described the application of this method to early type eclipsing stellar binaries, a transiting star-planet system with an early type host, and eclipsing double white dwarfs. 

Pursuing the measurement of the spin-orbit angle in eclipsing binaries with Sun-like components using photometry, as we described here, has a few advantages relative to other kind of systems and other methods. First, \ik is already collecting high precision photometry and is targeting primarily Sun-like stars. \ikp's data will eventually have a long time span, and will become publicly available. Second, the deep spectral absorption lines of late type stars allow obtaining high precision RVs, where the photometric results can, in principle, be confirmed. A similar confirmation will be more difficult for early type stars, due to the nature of their spectra, and even more so for double white dwarfs, since the eclipses are too short to allow sufficient exposure time.

%%%%%%%%%%%%%%%%%%%%%%%%%%%%%%%%%%%%%%%%%
\acknowledgments

% shporer
A.S. acknowledges support from NASA Grant Number NNX10AG02A.
% mazeh
T.M. acknowledges support from the Israel Science Foundation (grant No. 655/07).
% zucker
S.Z. acknowledges support from the Israel Science Foundation/Adler Foundation for Space Research grant no.~119/07.

%%%%%%%%%%%%%%%%%%%%%%%%%%%%%%%%%%%%%%%%%
 
%% Facilities
%{\it Facilities:} \facility{}

%\clearpage

%%%%%%%%%%%%%%%%%%%%%%%%%%%%%%%%%%%%%%%%%

\end{document}